\documentstyle[12pt,epsfig]{article}
\newcommand{\blankline}{\vskip .3cm}
\newcommand{\f}{\begin{equation}}
\newcommand{\ff}{\end{equation}}
\begin{document}
\centerline{\LARGE The present moment in quantum cosmology:}
\blankline
\centerline{\LARGE  Challenges to the arguments for the elimination of 
time}
\blankline
\blankline
\centerline{Lee Smolin${}^*$}
\blankline
\centerline{\it  Center for Gravitational Physics and Geometry}
\centerline{\it Department of Physics}
\centerline {\it The Pennsylvania State University}
\centerline{\it University Park, PA, USA 16802}
\centerline{and}
\centerline{\it The Blackett Laboratory}
\centerline{\it Imperial College of Science, Technology and Medicine}
\centerline{\it South Kensington, London SW7 2BZ, UK}
\blankline
\blankline
\centerline{August 30, 2000}
\blankline
\blankline\blankline
\blankline
\centerline{ABSTRACT}
Barbour,  Hawking, Misner and others have argued 
that time cannot play an essential role 
in the formulation of a quantum theory of cosmology.  Here we present 
three challenges to their arguments, taken from works and remarks by 
Kauffman, Markopoulou and Newman.  These can be seen to be based
on two principles: that every observable in a theory of cosmology should
be measurable by some observer inside the universe, and all mathematical
constructions necessary to the formulation of the theory should be
realizable in a finite time by a computer that fits inside the universe.
We also briefly discuss how a cosmological theory could be formulated 
so it is in agreement with these principles.

This is a slightly revised version of an essay published in 
Time and the Instant, Robin Durie (ed.) Manchester: Clinamen Press, 
2000

\vfill
${}^*$smolin@phys.psu.edu
\eject
\tableofcontents
\eject
\section{Introduction}

There is a general agreement that the notion of time is problematic 
in cosmological theories.  
During the last years discussions about time in cosmology have tended 
to focus on the question of whether time can be eliminated from
the fundamental statement of the laws and principles of the theory.
The argument that time can be eliminated has been put forward in its 
strongest form to date by Julian Barbour\cite{julian1,TEOT}.
I am convinced by Barbour's argument.  If a quantum cosmological theory 
can be formulated along the lines contemplated by many people in the 
field it follows, as he has argued, that time can be eliminated from 
the theory.  One of my first tasks here will be to outline the logic 
of the argument and show how that it does indeed follow from 
premises often assumed when talking about quantum cosmological 
theories.

However, before accepting the conclusion we should ask whether the 
premises are correct or not.  The main part of my argument will be 
that there are strong reasons to doubt the correctness of several of 
the key premises.  These  come from the unjustified 
reliance of the proposed quantum
theories of cosmology on mathematical structures that 
have no relevance for the representation of observations that can
be made by real observers inside the 
universe. Once these are eliminated it is no longer possible 
to make Barbour's argument. However it also becomes necessary to find 
a new way to formulate a fundamental theory of cosmology that does not 
allow the introduction of formal quantities which are observables in 
principle but unobservable in practice.  

Thus, the aim of my essay is to argue 
that the problem of time is genuine, but that its resolution
requires, not the elimination of time as a 
fundamental concept, but instead a reformulation 
of the basic mathematical framework we use for classical
and quantum theories of the universe. 

This reformulation is in any case
necessary in order to bring our theories into line 
with what working cosmologists actually do when they compare theory
and observation.  Real practice requires a theory that satisfies the
following two principles.

\begin{itemize}
    
    \item{}{\it Every quantity in a cosmological theory that is formally an 
    observable should in fact be measurable by some observer inside 
    the universe\cite{louis-holo,FM3,FM4,FM5,weakholo,weakstrong}.}
    
    \item{}{\it Every formal mathematical object in a cosmological theory 
    should be constructible in a finite amount of time 
    by a computer which is a subsystem of the actual 
    universe\cite{jim-worry,stulee1,stulee2,stu-invest}.}
    
\end{itemize}    

These principles have been enunciated before, in the cited references 
and elsewhere, but their implications have perhaps not been 
sufficiently explored.  What I aim to show here is that they are 
in contradiction with the 
argument for the elimination of time. Further, they imply that time 
will play an even more central role in the formulation of fundamental 
physical and cosmological theories. 

My argument will have four parts. In the first, I 
review the standard constructions used to formulate classical and
quantum cosmological theories and show that, in combination with the
two principles just quoted, they imply  
five postulates. In the 
second I sketch the basic argument why time is not a fundamental 
concept in theories which satisfy the five postulates. I believe that 
this account strengthens Barbour's claim that time is not a fundamental
concept in theories that satisfy these postulates.  In the third part
I will give three objections to the argument, which attack one or more
of the postulates.  In the  final part I mention some features
that a dynamical theory may have if it is to be consistent with  
these two principles.

In order to make the structure of the argument clear what follows is 
only a quick sketch of the argument. There is much more that could be 
said on the subject\footnote{Other implications of these issues
are explored in \cite{LOTC,3roads}.}.  
However my view is that this question is not one 
that can be settled by philosophical argument alone.  Were that 
possible the problem would already have been resolved.  The point of 
my argument is that the problem of time both requires and points to a 
chance in the structure of our physical theories.  What needs to be 
done next is to see if theories of this kind can be constructed and 
if they may also help to resolve other issues in physics such as 
quantum gravity and the 
foundations of quantum theory. I believe that the answer is likely 
yes, but this is something that cannot be argued for, it must be 
tried to see if it succeeds or fails.

Very little of my argument is original\footnote{The title is taken 
from a debate with Jaron Lanier,  held in Brooklyn, 
New York, April 3, 1999, under the auspices of Universitat 
Universalis.}.  Julian 
Barbour has greatly strengthened the argument for the elimination of 
time in cosmological theories, and the argument here was invented 
mainly in reaction to his recent papers\cite{julian1} and book\cite{TEOT}. I 
disagree with his conclusions but see no escape short of the kind 
contemplated here.  

The arguments against the elimination of time I 
present here are due to  
Stuart Kauffman\cite{stu-invest,stulee1},
Fotini Markopoulou\cite{FM3,FM4,FM5} and 
Ted Newman\cite{ted-personal}.  
Discussions with a number of other people,  
especially Jeremy Butterfield, Saint Clair Cemin,
Louis Crane, Fay Dowker, Chris Isham, Louis Kauffman, Jaron Lanier,
Seth Major,
Carlo Rovelli, Simon Saunders and Rafael Sorkin
have been crucial for my understanding of these 
issues.

Finally I must stress that both the positive and negative steps of my 
argument do not depend very much on the details of the cosmological 
theory under consideration. They apply as strongly in 
string theory as 
in quantum general relativity and they appear in all versions of these 
theories so far known.

\subsection{Some questions}

I would like to preface my argument with several basic questions
about cosmology.  

\begin{itemize}
    
    \item{}What is an observable in a cosmological theory, 
    based on general relativity?  What is the actual physical content 
    of the theory and how can we separate it out from its 
    mathematical presentation?  
    
    \item{}What will be an observable in 
    the ultimate theory of quantum cosmology?
    
    \item{}Is novelty possible?  Does general relativity allow the 
    existence of entities that do not exist at some time,  
    but exist at some other time (where a moment of time is 
    identified with a spacelike surface)?
    
    \item{}Will the ultimate quantum theory of cosmology allow the 
   possibility of novelty?
    
\end{itemize}    

The importance of the first question is that presently we can give a 
formal characterization of observables in general relativity, 
but we are actually not able 
to explicitly construct many examples of quantities that satisfy it.  
This stems from the fact that general relativity does not, 
as is often said,  identify the history of a 
physical universe with a manifold on which are defined a metric 
and perhaps other fields.  The correct statement is that the history 
of a universe is defined by an {\it equivalence class of manifolds 
and metrics under arbitrary diffeomorphisms\footnote{A diffeomorphism 
is in this context a map of a manifold to itself that preserves the 
notion of infinitely differentiable functions.  Thus, it moves the 
points around, but in a way that takes differentiable functions to 
differentiable functions. It thus preserves relationships between 
functions that can be described by coincidences of values at points}.}  

This is a key point, the significance of it is still often 
overlooked, in spite of the fact that it is far from 
new\footnote{The original argument for the identification of the 
physical spacetime with a diffeomorphism equivalence class of metrics 
is due to Einstein and is called the {\it hole argument.} It was 
strengthened by Dirac, Higgs, Bergman, DeWitt and others.  It is 
discussed in many places including 
\cite{stachel,carlo-time,LOTC,3roads}.}  
One major 
consequence of it is that there are no points in a physical spacetime.
A point is not a diffeomorphism invariant entity, for diffeomorphisms 
move the points around.  
There are hence 
no observables of the form of the value of some field at 
a given point of a manifold, $x$.  

If observables do not refer to fields measured at points what in the 
world do they refer to?  We have to begin with only the 
characterization that an observable must be a functional of the fields 
on a manifold, including the metric, which is invariant under the 
action of arbitrary diffeomorphisms.  

This is easy enough to state.  It is harder, but still possible, to 
describe a few observables in words. For example the spacetime volume 
is an observable for compact universes. So is the average over the 
spacetime, of any scalar function of the physical 
fields\footnote{Where the average is taken using the volume element 
defined by the spacetime metric.}. If the theory contains 
enough matter fields one can attempt to use the values of 
some of the matter fields as coordinates to locate points in generic 
solutions. Once points are labeled by fields, the argument goes, they 
have a physical meaning and one can then ask for the values of other 
fields at those points.

There are however several unsolved problems that make it doubtful that 
this is a satisfactory way to describe the observables of the theory. 
Past the first few simple ones such as spacetime volume we do 
not know whether the others are actually well defined on the whole 
space of solutions to Einstein's theory.  For example all attempts 
made so far to use the values of some physical fields as coordinates 
on the space of solutions fail because of the unruly behavior of the 
fields in generic solutions. As a result, we have control over only a 
handful of observables. But we need an infinite number of observables 
if we are to use them to distinguish and label the infinite number of 
solutions of the theory.

Finally, we must ask if any of the observables are actually measurable 
by observers who live inside the universe.  If they are not then we 
cannot use the the theory to actually explain or predict any feature 
of our universe that we may observe. 
If we cannot formulate a cosmological theory in 
terms that allow us to confront the theory with things we observe 
we are not doing science, we 
are just playing a kind of theological game and pretending that it is 
science.  And the worrying fact is that none of the quantities 
which we have control over as formal observables are in fact 
measurable by us.  We certainly have no way to measure the total 
spacetime volume of the universe or the spacetime average of some field. 

The reader may ask what relativists have been doing all these years, 
if we have no actual observables.  The answer is that most of what we 
know about general relativity comes from 
studying special solutions which have large symmetry groups. In these 
cases coordinates and observables can be defined using special tricks 
that depend on the symmetry.  These methods are not applicable to 
generic spacetimes, furthermore there is good reason to believe that 
many observables which are defined for generic spacetimes will break 
down at symmetric solutions.  

Thus, relativists have sidestepped the problem of defining observables 
for general relativity  and solved instead a much simpler problem, which 
is defining observables for fields moving on certain fixed 
backgrounds of high symmetry.  It is fair to say that the result is 
that we do not really understand the physical content of general 
relativity, what we understand is instead the physical content of a 
set of related theories in which fields and particles move on 
fixed backgrounds which are themselves very special solutions of the 
Einstein's equations. There is nothing wrong with this, so far as it 
goes, the problem is that this approach does not give us much 
information about the observables of the full theory, in which 
generic initial data evolves into a generic spacetime.  

The questions about novelty get their relevance from the fact that we 
observers in the real universe do genuinely observe novelty, in the 
sense that we observe things to happen that could  not have been
predicted on the basis of all the information that was, even in 
principle, available to us.  One source of novelty is that 
each year new stars and galaxies come into view that we could not, on 
the standard models of cosmology, 
have received light signals from before, due to the universe's finite 
age and the finiteness of the speed of light.  One may try to evade 
this argument in the context of an inflationary model, but this 
requires that we be able to predict the precise details of the light 
received from these distant galaxies on the basis only of the physics 
of their past during the inflationary era. But this is impossible in 
principle as the patterns of inhomogeneities that, according to the 
models of inflation, become the seeds for galaxy formation, are 
themselves seeded by quantum fluctuations in the vacuum state of a 
field during the inflationary era. Thus it is impossible 
in principle to predict the light that we will see next year arriving 
from a star presently too far to see, even assuming that inflation is 
correct.

A second source of novelty has to do with the fact that we live in a 
complex universe, so that we are constantly confronted with novel 
biological, sociological and cultural phenomena. It may seem   
that this has nothing to do with the problem of time in physics, 
but it is not completely obvious, given that a cosmological theory is 
supposed to allow us to make sense of all observed phenomena.  We will
see below that this kind of worry might under certain
circumstances indeed affect how we formulate
cosmological theories. 

In any case the first source of novelty is genuine and this is 
worrying enough. How are we 
to reconcile the fact that there is a necessarily unpredictable 
component to what we observe with the claim that time can be eliminated 
from our 
fundamental cosmological theories?  There may be an answer to this 
question, but this is an issue we will have to consider carefully 
before judging the claim that time can be eliminated from physics.

\section{The arguments for the elimination of time in cosmological 
theories}

Before giving the arguments for the elimination of time 
I must emphasize a crucial point, which is that 
they concern cosmological theories.  There is no problem of time in 
theories of isolated systems, embedded inside the universe 
or theories of systems with boundaries, 
at either finite or infinite distances. The reason is that if the 
system modeled is understood to include only part of the universe one 
has the possibility of referring to a clock in the part of the 
universe outside the system which is modeled by the theory.  This is 
generally what the $t$ in the equations of classical and quantum 
mechanics refers to\footnote{For example in asymptotically flat or 
Anti-DeSitter spacetime one can define time evolution with respect to a time 
coordinate at infinity.  There is a problem with how to continue this 
into the interior of the spacetime, but this is not the same as the 
problem which occurs in cosmological theory. It  can be resolved by an
appropriate choice of gauge, which means that, while it is a serious 
technical problem it is not a deep problem.}. The problem of time arises 
only in cosmological theories in which the whole universe is included 
in the degrees of freedom modeled in the theory, so that any clock, 
and any measuring instruments referred to in the interpretation of the 
theory must be part of the dynamical system which is modeled.

For the purpose of this discussion a universe is a closed 
system, which contains all that any part of it may interact with,
including any observers and observing instruments and any clocks used 
to measure time.
We will call any description of the physics of such a system a 
cosmological theory whether or not it is believed to be the actual 
universe, as its description faces the formal problems we are 
concerned with here.

There are two closely related arguments for the elimination of
time in cosmological theories, the first classical, the 
second its consequence for quantum theories of cosmology.

\subsection{The standard framework for classical cosmological theories}

The argument for the elimination of time in classical physics begins 
with the definition of a {\it configuration} of a universe.  
A configuration is
a possible state or situation that the universe can have at a given 
moment of time.  The arena in which the argument takes place is 
the {\it configuration space} of the universe
denoted $\cal C$. This is defined to be the space of all possible 
{\it configurations} that the universe can be in at a given 
moment of time. 

For a single particle restricted to be in a room the 
configuration space, ${\cal C}_{room}$ is defined to be the three 
dimensional space 
contained by the room; each point in the room is a place the particle 
might be and hence a possible configuration of the system.  
For a system of $N$ particles in the room the configuration space
is ${\cal C}_{room}^N$.

In classical physics it is assumed that universes have 
{\it histories}, which are described as curves
$x^a(s):R \rightarrow {\cal C}$ in the configuration space. 
$s\in {\bf R}$ is called the {\it time parameter}.  Part of the 
question we are concerned with is the role of these histories in the 
description of the universe and the relationship between time 
parameters and time as measured by a clock carried by an observer 
inside the universe\footnote{A rather different question is 
the relationship between the time parameters and our experience of 
the flow of time.   For the purposes of this paper we 
will be able to ignore this question, as the answers to the 
questions that will be considered will change considerably the context 
for its consideration..}.

In the introduction I enunciated two principles that a cosmological 
theory must satisfy to be relevant to observations made by real 
observers inside the universe. In combination with the formalism 
of classical cosmological theories they lead to five postulates. 
We begin with the first three.

\begin{itemize}
    
    \item{}{\bf Postulate A: Constructibility of the configuration space.}
    It is possible by a finite mathematical procedure for observers 
    inside a universe to construct its configuration space ${\cal C}$.
    
    \item{}{\bf Postulate B: Deterministic evolution}
    There is a law 
    for the evolution of universes which, given a position 
    $x^a(0)=q^a \in {\cal C}$ and 
    a velocity vector, $v^a\in T_q$ at $q^a$, picks a trajectory 
    $x^a (s)$which is unique up to redefinitions of the time parameter
    $s$.  The law is given by an action principle $S[x^a(s),v^a(s)]$.
    
\end{itemize}

The third postulate should tell us something about the observability
of the configuration space. Ideally we should require something like,

\begin{itemize}
    
    \item{}{\bf Idealized  Postulate C: Observability of the configuration 
    space}  It is possible for observers inside a universe to make 
    measurements which are sufficient to determine which trajectory 
    $x^a (s)$ describes the history of our universe.  
    
\end{itemize}    

However this is probably too strong, as real observers are much smaller
than the universe and hence are unlikely to be able to gather and 
store enough information to determine its precise history. However, 
at the same time we want our observers to be able to measure enough
observables to make some predictions as to the future values of some 
observables.   The solution to this was understood a long time ago
by the people working on the consistent histories approach to quantum
theory.  They proposed a notion of coarse grained histories.  This 
will not, however, serve for us here, because that expresses the 
cosmological theory irreducibly in terms of histories, hence it cannot
be a context in which we can run the argument for the elimination of 
time. What is required to do that is some notion of a coarse grained
configuration space, $\tilde{\cal C}$ whose observables correspond 
to suitable averages of 
observables of the physical configuration space. From the evolution law 
    specified in Postulate B we can then derive a law for the 
    evolution of probability densities on the coarse grained
    configuration space.   However, for the usual practice of 
    classical cosmology we require more: it must be the case that
    we can derive an approximate deterministic evolution law on the
    coarse grained configuration space. An example is the reduction
    to a minisuperspace model, by eliminating the spatial dependence
    of the spacetime metric, which then satisfies a reduction of the
    classical Einstein equations with a finite number of degrees of 
    freedom.  

\begin{itemize}
    
    \item{}{\bf  Postulate C: Observability of a coarse graining of 
   the configuration  space}.It  must be possible to define, by a 
   suitable averaging procedure from the configuration space, 
   ${\cal C}$, a coarse grained
   configuration space, $ \tilde{\cal C}$, together with an evolution
   law, such that an observer inside the universe can make measurements
   sufficient to determine the trajectory in $ \tilde{\cal C}$.
    
\end{itemize}

We may note that each postulate is necessary for cosmology to be 
treated according to the usual methods of classical physics.  If
{\bf A} is not satisfied than we observes inside a universe cannot 
describe it according to the methods of classical physics.  If
{\bf B} is not satisfied than we cannot use the theory to make predictions
of the future or retrodictions of the past. If 
{\bf C} is not satisfied than we cannot compare those predictions and 
retrodictions with observations.

Given the definition we have given of a universe as a closed system, 
which contains all that any part of it interacts with, there is no 
role whatsoever in the notion of {\it an observer outside the universe}.
Any recourse to such an observer in discussions of cosmological 
theories represents an implicit admission that the theory under 
discussion is not a proper framework for doing cosmology. This is
the reason we insist on the two principles enunciated in the 
introduction.  

All modern approaches to cosmology take on board the 
principle that observations of the configuration of the universe are 
relational in the sense that they refer to coincidences in the values 
of variables observable from inside the universe.  This principle was 
first enunciate by Leibniz and the exact sense in which it is 
satisfied by general relativity is explained in writings of 
Barbour\cite{julian1,TEOT}.  We may express this as a fourth
postulate

\begin{itemize}
    
    \item{}{\bf Postulate D: {\cal C} is a relative configuration space.}
    This means that a point $x^a \in {\cal C}$ is completely 
    determinable by measurements made by observers inside the universe.
    
\end{itemize}    
    
This has the following 
important consequence:{\it Given that any two distinct points $x^a$ and $y^a$ of
${\cal C}$ must refer to  configurations that can 
be distinguished by observers inside the universe
there can be on $\cal C$ no global symmetry such as 
Euclidean invariance and no local gauge invariance such as 
diffeomorphism invariance.}

The exact form of the relative configuration 
space\footnote{Motion on such relative configurations spaces
was first studied by Barbour and Bertotti and Barbour\cite{BB}.} depends on the 
content of the theory.  For a system of $N$ particles in $d$ 
dimensional space, the
relative configuration space may be constructed as a quotient,
\f
{\cal C}_{rel} = {R^{dN} \over {\bf Euclid}}
\ff
where $\bf Euclid$ is the Euclidean group of rotations and translations in
$R^d$.

For general relativity and related theories such as supergravity 
\f
{\cal C}_{gr} = {\mbox{metrics and fields on} \Sigma  \over Diff(\Sigma )}
\ff
where $Diff(\Sigma )$ is the group of diffeomorphisms on a compact 
manifold $\Sigma$, which is taken to represent ``a spatial slice of the 
universe."

An important point, to which we will return later, is that the 
relative configuration spaces are, at least in the examples studied
so far, defined as quotients of a well 
defined space by the action of a group.

Nor can there be any notion of a clock which is something other than a 
degree of freedom measurable by observers inside the universe.  Since 
all observables are assumed to be relational, and hence to measure 
coincidences in values of measurable quantities, there can be no 
reason why one clock can be preferred over another, so long as the 
values of the two of them can be related to each other uniquely.  The 
result is our final postulate.  

\begin{itemize}
    
      \item{}{\bf Postulate E: Reparameterization invariance}  Two
    trajectories $x^a (s)$ and $x^a (s')$ which differ only be a 
    redefinition of the time parameter, $s'=f(s)$ are deemed to 
    describe the same physical history of the universe.  This implies 
    that the action $S[x^a(s),v^a(s)]$ is invariant under these 
    redefinitions.
    
\end{itemize}

\subsection{The classical argument for the elimination of time}

The five axioms we have given define a classical cosmological theory. 
They rule out the existence or relevance of any clocks outside the 
system.  Postulate {\bf E} also rules out any absolute internal time,
in that all time parameters appear, at least at first, to be on an 
equal footing.  We must then wonder what the proper notion of time
is in such a theory and, in particular, if it contains any concept of 
time that can be connected with our observations.

I will claim, following arguments by Barbour and others, that the
theory allows no fundamental notion of time. By this we mean that the 
theory can be formulated in such a way that no reference is made to a 
time parameter.  This does not exclude the introduction of parameters 
which have, at least in particular solutions, some of the properties 
of time in ordinary theories.  Often these are degrees of freedom 
which behave in some regimes like physical systems we call clocks.
We may thus call these clock variables.
Typically these behave as would be 
expected in some regions of the space of solutions, but not in all.
For these reasons, they may provide an approximate or effective notion 
of time in some domains in some solutions.  But if the theory can be 
formulated without any notion of time then there is no guarantee that 
these approximate clock variables measure some more fundamental 
quantity which deserves the name of a universal time for the theory as 
a whole.

The basic argument for the elimination of time is the following.  Given {\bf B}, 
any pair $(x^a (s),v^a(s))$ determines
a unique history $x^a(s)$.  Conversely, given any history determined 
by the laws, and an arbitrary choice of initial time labeled by 
$s=s_0$, the whole history of the universe $x^a (s)$ is determined by its 
initial data
$(x^a (s_o),v^a(s_0))$.   
But any observable $\cal O$
measurable by an observer inside 
the universe must be a function of the pair $(x^a (s),v^a(s))$. 
As a result any such function is actually determined completely by the 
initial data $(x^a (s_o),v^a(s_0))$. 

But no physical observable can depend on the actual value of the time
parameter, $s$ because, by {\bf E}, all observables must be invariant
under rescaling of the time parameter.   This means that time can
only be measured in terms of coincidences between the values of
the $(x^a (s),v^a(s))$ which do not depend explicitly on the specification of
the actual parameter $s$ at which that coincidence occurred.  (An 
example of such a coincidence is to ask what the value of $x^2$ is
when $x^1$ equals $17$.)  But such observables must be, by
determinism, functions only of the trajectory.  
This means that any such observable must have the 
same value on any two points of the same trajectory. 

This conclusion may seem counterintuitive when first encountered,
if so it may help to go over.  
the argument a bit more carefully.  
The point is that the time parameter $s$ is not measurable by any 
observer.  Because it can be changed without consequence to the 
physical history, its only role is as a mathematical device to label
the different points on a trajectory.  Since that labeling is 
arbitrary, it cannot correspond to anything that an observer inside 
the universe can measure, in particular it cannot correspond to the 
reading on a physical clock.  

Something which is observable then must be expressed as a correlation 
between different functions of the pairs $(x^a(s),v^a(s))$ that is
independent of the parameterization $s$.  For example it may be a 
correlation between the reading of one dial of an instrument and 
another.  These correlations are independent of the parameterization 
and can be used to define a physical notion of time which is 
observable by an observer inside the system. But as it is independent 
of the parameterization, such a correlation must be determined by the 
$(x^a(s_0),v^a(s_0))$ for any arbitrarily chosen parameter value
$s_0$.  But since $s_0$ is arbitrary this means the observable is
actually a function of the trajectory. If it is to be expressed
in terms of the $(x^a (s),v^a(s))$ it must be in a way that is
constant along each trajectory.

This is a key step of the argument, for the complete version I refer the 
reader to papers by Barbour\cite{julian1,TEOT} and
Rovelli\cite{carlo-time}.  

Mathematically this has the following consequence. Because each  
pair $(x^a (s),v^a(s))$ determines a unique history, the evolution 
along each trajectory defines a one parameter group of diffeomorphisms
$E$ of $\cal C$, each member of which takes each point in $\cal C$ to 
another point on the same history.   The argument just given says 
that any physical observable $\cal O$ must be invariant under the action of
$E$.  As a consequence one can 
take the quotient by the action of $E$.  To do this it is convenient
to first go to the phase space of the system, $\Gamma$, which is
defined in simple theories\footnote{In all theories the phase
space is defined by position-momentum pairs, only in simple
theories is the momentum proportional to the velocity. This 
technical point is not relevant to the argument being made here.}
to be the $2N$ dimensional space of 
pairs $(x^a (s),v^a(s))$.
The reason is that the evolution on the phase space is first 
order in time. This means that
only one trajectory can go through any point on the phase space.
One can then define the quotient,
\f
{\cal U} = {\Gamma \over E} .
\ff
This defines $\cal U$, the space of physically distinct histories
of the universe. Since all observables must be invariant
under the action of $E$, it follows that 
any observable $\cal O$ is actually a function on $\cal U$. 

The final form of the theory is then that the possible universes are described 
as points of $\cal U$.  All observables are functions on $\cal U$.  
The notions of time and trajectory  
have disappeared from the final form of the theory.

The definition of the quotient may involve some technical work, but is 
assumed to be always possible. One way to do it is to defined a
gauge condition, which is an equation for a surface in $\Gamma$ that
intersects each history exactly once.  One may then identify 
$\cal U$ with this surface.  Physically, one way to do this is to 
define a physical time parameter in terms of some observable quantity
$T(x^a) $ which is a function on $\Gamma$ which has the property that 
$(x^a)=0$ defines such a surface.  

The result is that time has been eliminated completely from the 
theory. One does physics by determining which trajectory one is on and 
then determining the value of the observables, expressed as 
correlations between quantities measurable by local observables.  
Some of these quantities may be interpretable as readings 
on devices we call clocks, in which case we can recover at some level 
a notion of time, defined just as the readings on a device called a 
clock. But there is no reason we must interpret the observables in 
terms of such clock variables.  Time may or may not be a useful 
construct, good for some level of description.  But it has no 
fundamental role in the theory.

\subsection{The argument for the elimination of time in quantum cosmology}

We now turn to discussion of the quantum theory.  We will take the 
conventional  view that a quantum theory of cosmology should be a 
quantization of a classical theory of cosmology.  This is of course 
unlikely to be true, as the quantum theory is assumed to be the 
fundamental theory and the classical theory should be derived from it 
by some suitable approximation procedure. But we will adopt the 
conventional view here as it is the context in which the argument for 
the elimination of time in quantum cosmology follows.  As before I 
will only sketch the argument, details can be found in 
\cite{julian1,TEOT,carlo-time}.

There are several different approaches to turning a classical 
cosmological theory into a quantum theory. We describe only one here, 
the logic in the other approaches are similar.  This is the approach
based on the Wheeler-DeWitt, or {\it hamiltonian constraint equation.}
This approach arises in the context of hamiltonian quantization of
theories which satisfy postulate, {\bf E}, invariance under arbitrary
redefinitions of the time parameter.  To make the argument as 
transparent as possible to those unfamiliar with it, and to avoid 
boring those who are, I skip the steps of the construction and 
exhibit only the result.

In this approach, a quantum theory of cosmology depends also on five postulates.
Postulate {\bf A} is taken over directly
from the postulates of classical cosmology. 
The postulates {\bf B} and {\bf C} are replaced by

\begin{itemize}
    
    \item{}{\bf Postulate $B^\prime$: Existence of the wavefunctional of the 
    universe.} A quantum state of the universe is defined to be a 
    normalizable complex functional $\Psi$ on
    $\cal C$.  This means that there exists a measure $d\mu$ on $\cal C$ 
    such that 
    \f
    1=\int_{\cal C} d\mu |\Psi|^2 .
    \ff
     The normalizable states define a 
     Hilbert space on which $\mu$ defines an inner product according to
     \f
     <\Phi|\Psi> = \int_{\cal C} d\mu \Phi^*\Psi
     \ff
     Observables must then be represented as self-adjoint operators in this 
     inner product space.  As in ordinary quantum mechanics there 
     must be complete commuting sets of observables whose spectrum 
     determine the quantum state uniquely.  This gives rise to
     
     \item{}{\bf Postulate $C^\prime$: Observability of the configuration 
    space}  It is possible for observers inside a universe to make 
    measurements which are sufficient to determine a density
    matrix $\rho$ which has sufficient information to determine, in a 
    suitable classical limit, a trajectory in the coarse grained
    congifuration space $\tilde{\cal C}$.

\end{itemize}

Postulate {\bf D} is either taken over directly, or replaced by a
set of first order functional differential equations called the 
diffeomorphism constraints $D(v)$, which depend on an arbitrary 
vector field $v^a$ on $\Sigma$, the spatial manifold.  This expresses 
the requirement that the quantum states are valued only on relative 
configurations which are in this case diffeomorphism equivalence 
classes of metrics on $\Sigma$. In the latter
case we have

\begin{itemize}
    
    \item{}{\bf Postulate $D^\prime$: Diffeomorphism constraints.}  The 
    wavefunctionals satisfy also
    \f
    D(v)\Psi =0
    \ff
    for all vector fields $v^a$ on $\Sigma$. 
    
\end{itemize}    

Finally, the crucial postulate {\bf E} that expresses time 
reparameterization invariance is replaced by 

\begin{itemize}
    
    \item{}{\bf Postulate $E^\prime$: Solution to the Hamiltonian constraint.} 
    There is a second
    order functional differential operator, $ H$ on $\cal C$
    such that
    
   \f
   { H}\Psi =0
   \label{hamc}
   \ff
\end{itemize}    

Of course many technical problems arise in the realization of these 
conditions.  An important motive for taking them seriously is that 
they can in fact be realized in quantum general relativity in $3+1$ 
dimensions, coupled to arbitrary matter 
fields\cite{loops}\footnote{This formulation, called loop quantum 
gravity, comes from using for the configuration space the 
diffeomorphism and gauge equivalence class of a certain connection
on $\Sigma$, rather than the metric.  This change makes it possible to
obtain precise results, but does not affect the conceptual arguments
under discussion here.}.

 There are many interesting issues concerning the 
construction and interpretation of such observables.  We do not need 
to go further into this discussion than to note that as in the 
classical case time plays no role in the actual 
formulation of the theory. 
As many have remarked, the Hamiltonian constraint equation,
(\ref{hamc}) contains the dynamics of the theory, but there is
no $d/dt$ on its right hand side. This is because any $t$ not
contained in the wavefunctional would be a non-dynamical time, 
disconnected from the dynamical system described by the 
wavefunctional.  But the basic postulates of the theory tell us there 
can be no such external time parameter.  As in the classical case,
time parameters may be introduced, but the physical observables
cannot depend on the actual value of any time parameter.
Hence the right hand side of the equation
is $0$. Rather than expressing the dependence of the state on an 
external time, as in the Schroedinger equation, the constraint 
expresses only the fact that the quantum state has no dependence on 
such an external time.

A full specification of the theory is given by the choice of 
configuration space, Hamiltonian (and possibly other) constraints.
There is no place for a time parameter, time has truly been eliminated 
in the theory.  Nor does time necessarily show up in the form of any 
solutions, many solutions are known \cite{loops,tedlee,carlolee} that 
have nothing like a time parameter.

There may be approximate notions of time 
which arise from properties of the solutions or observables, but no 
notion of time is needed to construct the theory.  If a theory 
formulated along these lines is correct, time has 
disappeared from the fundamental notions needed to describe the 
physical universe.

\section{Challenges to the argument for the elimination of time}

We now turn to several challenges which have been made recently to 
the argument just sketched. To my knowledge these are new and in my 
opinion they deserve careful consideration.  If they are right then 
not only is time still a necessary concept, we will have to find a 
different way to frame dynamical laws.  The success  of these challenges 
then cannot depend only on whether or not they point up a failure of the 
argument for the elimination of time. Their contribution must be at 
least equally positive as negative, they must point us towards the 
invention of a new dynamical framework in which time plays a 
different, and more essential, role than at present.  

The arguments for the elimination of time as a concept necessary for the 
expression of the fundamental laws of physics follows from the five
postulates mentioned above.  If they all hold then the argument
goes through.  At a purely formal level, they do seem to hold for 
quantum general relativity in $3+1$ and more dimensions, coupled to 
any kind of matter.  The progress of the last 16 years using the
Sen-Ashtekar formalism\cite{sen,abhay} and 
loop quantum gravity\cite{loops} strengthens the 
argument as it leads to the explicit construction of solutions to the
Hamiltonian constraints\cite{tedlee,carlolee}. 
Supersymmetry seems to make no difference, nor is 
there any reason to believe that they are not satisfied for string 
theory, even given that we do not know the framework of string theory 
at the background independent or non-perturbative 
level\footnote{However, it is clear that time can be eliminated in
the proposals that have so far been made for a background
independent form of {\cal M} theory\cite{horova,1616}.}.  

If the argument goes through then there seems little alternative to 
agreeing with the point of view expressed eloquently by Julian Barbour 
in \cite{TEOT}.  This is the end of the concept of time in 
fundamental physics.

There seems 
only one real hope of evading the argument, which is that there is in 
fact no way to realize all five principles on which the 
argument depends consistently in a single theory.  Is this possible?  It is 
certainly the case that there are consistent and completely worked out
{\it models} of quantum 
theories of cosmology which have only a few 
degrees of freedom.  These include quantum gravity in $1+1$ and $2+1$ 
dimensions as well as some models of very symmetric universes. However 
these only satisfy postulates {\bf B, D} and {\bf E}. They fail
to satisfy the other postulates because they are too 
simple to contain subsystems complex enough to be called an observer.
Further, despite all the progress made on the quantization of general 
relativity, supergravity and related 
theories\cite{loops,tedlee,carlolee}, the explicit construction of
an infinite number of physical states has {\it not} been followed by
the construction of an infinite number of physical observables.
It is then an open question whether there are any theories which 
satisfy all of the postulates.  

\subsection{A first challenge: are there observables without time?}

Postulate {\bf C} requires the construction of a sufficient number of 
observables of the theory to distinguish the solutions from each 
other.  As we are dealing with a theory with an infinite number of 
degrees of freedom this means we must have an infinite number of 
observables.  In the introduction we discussed some of the issues 
involved with the construction of observables in general relativity.
To sharpen this discussion we may distinguish two possible approaches 
to the construction of observables in classical and quantum theories 
of gravity

\begin{itemize}
    
    \item{}{\bf Causal observables}. These are instructions for the 
    identification of observables that make explicit reference to the 
    causal structure of a classical or quantum spacetime.  Since the 
    causal structure is a diffeomorphism invariant of the metric, such 
    an observable may be diffeomorphism invariant by construction.  
    Examples are known which are of the following form: Identify a 
    particular localized system as a local reference system and 
    identify one of its degrees of freedom as a clock. Define 
    observables in terms of the values of other local degrees of 
    freedom coincident with the clock variable taking on particular 
    values.  These correspond to actual observations that could be 
    made in a spacetime, which, using the causal structure, give 
    information about the spacetime to the past of the event when 
    that local clock variable had a particular value.
    
    Such an observable can be constructed explicitly in a histories 
    formulation of the theory without solving any additional 
    conditions.   They can be in principle be directly 
    implemented in a path integral 
    formulation of the theory based on summing over Lorentzian 
    histories.   
    
    \item{}{\bf Hamiltonian constraint observables.}These are 
    observables which are constructed according to the rules of the 
    hamiltonian formulation for systems with time reparameterization 
    invariance. They must do at least one of the following things,
    i) have vanishing Poisson bracket 
    with the classical hamiltonian constraint, ii) be expressed directly as a 
    functional on the reduced phase space which is the constraint 
    surface mod gauge transformations. iii) commute with the 
    quantum hamiltonian constraint.

\end{itemize}    
 
Observables of the first kind make explicit use of the causal 
structure and hence use time in an essential way in their 
construction.  If we only work with these observables then we have 
not eliminated time from the theory.  Furthermore, there are no 
obstacles to defining and working with such observables as no 
equations need to be solved.  They are diffeomorphism invariant by 
construction.  

On the other hand, if we eliminate time from the theory, as sketched 
above, by either reducing the classical theory to its reduced phase or 
configuration space or constructing a quantum theory from solutions to 
the Hamiltonian constraint, then we have 
available only the second class of observables.
We then conclude that Postulate B requires that we be able to 
construct an infinite number of observables of the second kind.  
Further, all observables of the first kind must be reducible to 
observables of the second kind.

The problem is that Hamiltonian constraint observables are extremely 
difficult to construct in real field theories of gravitation. There are formal 
proposals for how to construct such observables, which have been 
implemented in toy models. But these toy models are too simple and do
not have local degrees of freedom that could be identified with 
fields measured by local observables inside the spacetime, or with 
such observables themselves.  
No observables have ever been constructed through either of the three 
methods mentioned which are local in the sense that they correspond 
to what an observer inside a relativistic spacetime would see. 

The problem that arises in method i) is that to give such an 
observable explicitly as a functional on the phase space of the theory 
requires explicitly inverting the equations of motion of the theory over the 
whole space of possible initial data.   This cannot be done for
other than integrable systems.  Similarly,  method ii)
has not been implemented because no one has a proposal for how to 
actually construct the reduced phase space. This also would involve 
an inversion of the equations of motion of the theory.
The problem with method iii)
is that while we have been fortunate enough to find an 
infinite number of quantum states which are 
exact solutions to the Hamiltonian constraints of 
theories of gravity\cite{tedlee,carlolee}, finding operators which commute 
with the 
constraints has proved to be much harder, and only a few, rather 
trivial such operators have been found.  No one has even proposed a
practically implementable  strategy 
for how to construct operators that both commute with the quantum 
Hamiltonian constraint and refers to local observations.  

So whether or not there exist in principle observables of the second 
kind, there are no known methods to construct them for realistic 
theories.

\subsection{Newman's worry: the implications of chaos}

Ted Newman\cite{ted-personal} and others have raised a further 
argument against the 
existence of observables of the second kind, constructed through 
either method i) or ii) which is that the fact 
that gravitational systems have chaotic behavior may in principle
prevent their construction.  It is well known that the motion of three 
bodies under their mutual gravitational attractions 
is chaotic in Newtonian mechanics.  It is then hard to believe that 
the typical solution of general relativity is not also chaotic. It is 
also known that the generic cosmological model with a finite number 
of degrees of freedom, called the Bianchi IX model, is chaotic, at 
least if a physically realistic notion of time is used to in the 
criteria for chaos. If we do not yet have a proof of that the Einstein 
equations are chaotic it is mainly because of the problem of finding a 
definition of chaos suitable for the application to a field theory of 
gravity.

But if the equations are chaotic it means that 
physical observables are not going to be representable by smooth functionals on 
the phase space.  The reason is that they must be constant under the 
trajectories defined by the hamiltonian constraints, as this must be 
satisfied by any physical, diffeomorphism invariant observable. But 
if the system is chaotic than any trajectory passes through any open 
set in the phase space.  This means that non-trivial physical observables will 
take an infinite number of values in any open set. It is easy to 
construct examples of such observables which are not continuous 
anywhere inside at least certain open sets.

It will be difficult both to construct such observables explicitly
and to evaluate them. It may also not be possible to define an 
algebra of such observables, as this requires taking derivatives to 
form the Poisson bracket.   But if such observables do not have an 
algebra they cannot be represented in the quantum theory, as the 
quantum theory is generally defined in terms of a relationship between 
the algebra of quantum operators and the Poisson algebra of classical 
observables.

Newman's worry turns what may seem at first only a practical problem 
into a problem of principle.  It may be not only beyond our power to 
construct observables for a formulation of cosmology in which time has 
been eliminated, it may be impossible in principle. If this is the 
case then the argument for the elimination of time fails, 
for a theory without observables that 
can be made to correspond to what we real observers can measure is 
not a physical theory, it is just a formal structure with no possible 
relation to the real world.

\subsection{Markopoulou's argument: the configuration of the universe 
is not observable}

In a recent article Markopoulou\cite{FM3} pointed out that postulate C, 
observability of the configuration space is not likely to be satisfied 
by any quantum theory of gravity whose classical limit is general 
relativity. The reason is that in generic cosmological solutions to 
classical general relativity containing dust and radiation, with
spacetime topology of the form $\Sigma \times R$,
with $\Sigma$ either $R^3$ or $S^3$, the
backwards light cones of typical events do not contain the full Cauchy 
surface $\Sigma$. This means that Postulate C fails in these examples.

It is important to stress that the problem is not just the inability
to measure enough observers to determine a trajectory in the fine
grained configuration space ${\cal C}$.  The problem is that as
we see only a portion of every Cauchy surface, we cannot measure
enough information to determine a history within any coarse
grained configuration space.  As George Ellis has pointed 
out\cite{ellis} we cannot make use of a coarse grained cosmological
model such as a minisuperspace model unless we supplement our 
assumptions with the {\it assumption} that the portion of the 
universe we see at any time is a typical region of the whole universe 
at that time.  (Note that this also assumes sufficient homogeneity 
that it makes sense to talk of a simple cosmological time function.)
This assumption may be dressed up into a principle, the so-called
cosmological principle, but there can be no getting away from the
fact that it is an assumption that does not, and very possible cannot have strong
support from observation.  Furthermore, the assumption is strictly 
false in many plausible cosmological models such as inflationary models.

Once one accepts the possibility that the cosmological principle may
be false, Markopoulou's argument has force, because it implies that
any internal observer is unable to gain from observation enough 
information to carry out a coarse graining which can satisfy 
either the classical or quantum version of Postulate C.

It is possible to avoid this conclusion in several ways. The first is 
to have a ''small universe" in which the spatial topology is more
complicated.  Presently this appears to be ruled out by observation.
The second is to weaken the requirement that the matter be described 
by dust and radiation.  By doing so we can allow inflation to have 
occurred in our past.  There are, however, two problems with saving 
the observability of the universe by inflation. The first is that 
without fine tuning inflation predicts that the universe is spatially 
flat.  In this case we certainly do not see a whole Cauchy surface in 
our past.  The second is that according to inflationary models the 
whole Cauchy surface would be seen only during the inflationary
era when the quantum 
fields are close to their ground state. The inhomogeneities that drive 
structure formation in our universe are hypothesized to arise only at 
the end of the inflationary period, in a ``quantum to classical 
transition" in which quantum fluctuations in the vacuum state are 
converted to classical fluctuations in the matter and geometry.  This 
transition is believed to be akin to a measurement of the quantum 
state of the fields. The result is that the information necessary to 
determine the classical geometry of the universe is nevertheless not 
available in the backwards light cone of any one observer.

More could be said about these discussions, but the conclusion is
that Postulate C is likely not satisfied in our universe.

We are then faced with the following choice: Give up postulate C and 
accept a classical or quantum theory of cosmology which is formulated 
in terms of quantities that are not observable or attempt to modify the 
theory so 
as to formulate it only in terms of quantities observable by real 
observers inside the universe.  In \cite{FM2} Markopoulou describes a 
framework in which general relativity and other spacetime theories 
may be reformulated completely so that the only observables involve 
information available to internal observers by means of information 
reaching them from their backwards light cones.  

The potential power of Markopoulou's argument rests on the following 
simple observation: in classical general relativity the causal 
structure plays a significant role in delineating what is observable.  
To begin with Markopoulou points out that if we insist on the 
principle that any observable of a cosmological theory must be in 
fact observable by an observer inside the universe than it follows 
that our theory must be pluralistic in the particular sense that 
different observers have access to different information.  However 
this is a structured pluralism because the causal structure implies 
certain logical relations amongst the observations made by different 
observers.  Markopoulou then points out that the pluralistic logic 
required to define what is observable in general relativity is 
closely related to logics studied by mathematicians under the name of 
topos theory.  The basic rules which relate observations made by 
different observers follow from the causal structure and the 
requirement that whenever two observers receive information from the 
same event they will agree on the truth value given to any 
propositions about that event.  

Markopoulou's argument then formalizes the worry about novelty
I mentioned in the introduction.  The key point is that any 
given observer, located in some finite region of space and time, only 
receives information from a proper subset of the events in the history 
of the universe.   Furthermore, since no observer sees a whole Cauchy 
surface, they have no way to predict what information will be 
received from systems of the universe they will come into causal 
contact with at a later time.  As a result, the logic of observables 
in a cosmological theory must not satisfy the law of the excluded 
middle. No single observer can assign the values {\bf TRUE} or {\bf FALSE} 
to all 
of the propositions which describe the history of the universe.  There 
must be other truth values such as, {\bf cannot yet be determined by 
this observer}.  
In turns out that one has to be able to describe 
different degrees of ignorance, so that there is 
actually a whole algebra of possible truth values besides true 
or false\cite{FM3,FM4,FM5}.

The result is that the logic of observables in a relativistic 
cosmological theory must be both non-boolean and pluralistic.  For any 
single proposition, different observers will be able to assign different 
truth values.  Whenever two are able to assign {\bf TRUE} or 
{\bf FALSE} they must 
agree.  But they need not agree if one or both are only able to assign 
truth values which indicate some amount of inability to determine.

So far we have been discussing only the classical theory.  There are 
also implications for the quantum theory which Markopoulou develops in 
further papers\cite{FM4,FM5}.  The point is that the possible truth 
values of the classical theory must appear in the spectrum of 
projection operators in the quantum theory.  But, she points out,  
one cannot realize 
such a non-Boolean pluralistic logic in terms of the spectrum of 
operators on a single Hilbert space.  Consequently, a quantum theory 
of cosmology whose classical limit reproduces the algebra of 
observables of the classical theory explored by Markopoulou cannot
be based on a single Hilbert space.  As she describes in \cite{FM3,FM4},
it can, however, be constructed from a presheaf of Hilbert spaces, 
which is a structure in which there is a Hilbert space assigned to 
every event in a causal set.  

The result is a new framework for quantum cosmology in which the 
causal structure, and hence time, plays a fundamental role in the 
formulation of theory.  The basic implication of Markopoulou's work 
for the present argument is then that to the extent that the causal 
structure plays a role in the identification of observable quantities 
in general relativity, it must also play an essential role in the 
construction of the corresponding quantum theory.  If we are then 
restricted to the first kind of observables, which are causal 
observables, than time cannot be eliminated from quantum cosmology. 
 
\subsection{Kauffman's argument: the configuration space cannot be 
described in advance of the evolution of the system}

We come now to a challenge to the argument that attacks the first 
postulate, which is the constructibility of the 
configuration space\footnote{This 
kind of worry goes back at least to a paper by James Hartle in which he 
points out that the impossibility of classifying four manifolds is a 
problem for the Euclidean path integral formulation of quantum 
cosmology\cite{jim-worry}.}.  This challenge was inspired by two
very interesting questions raised by Stuart Kauffman in the course of
recent work on theoretical biology and economics\cite{stu-invest}: 

\begin{itemize}
    
\item{}{\it Is it possible
that the configuration 
spaces relevant for mathematical biology, ecology or
economics cannot be constructed by any finite mathematical procedure?}

\item{}{\it Even if the answer to the first is in principle no, is it 
possible that the construction of the relevant configuration spaces are so 
computationally intensive that they could not be carried out by any 
subsystem of the system in question?}   

\end{itemize}

There are good reasons to suspect that the answer to at least one
of these questions is yes when we are
dealing with such potentially large and complex configuration spaces
such as the space of all possible 
phenotypes (as opposed to genotypes) for biological species, the 
space of all properties that might be acted on by natural selection,
the space of all biological niches,
or the space of possible kinds of businesses or ways of earning a 
living.

We refer the reader to \cite{stu-invest} for discussions of
the implications of this possibility for theories of biology and 
economics.  For the present purposes we are interested to consider the 
analogous questions\cite{stulee1}:

\begin{itemize}
    
    \item{}{\it Is it possible that there is no finite procedure by means 
    of which the configuration space of general relativity or some 
    other cosmological theory may be 
    constructed?}
    
    \item{}{\it Even if the answer is no is it possible that the 
    computation that would be required to carry out the construction 
    of the configuration space is so large that it could not be 
    completed by any physical computer that existed inside the 
    universe?}
    
\end{itemize}    

The possibility that the answer to one or both questions is yes arises 
from the fact that, as pointed out in section 2, the physical 
configuration spaces relevant for cosmological theories are quotients
of infinite dimensional spaces by the action of infinite dimensional 
groups. For the case of general relativity in $3+1$ or more 
dimensions the physical configuration space is defined to be the 
quotient of the space of metrics on some fixed compact manifold, 
$\Sigma$, by the diffeomorphism group of $\Sigma$.  No closed form 
representation of the quotient is known, even in the simplest case
in which $\Sigma = S^D$.  The space of metrics on a manifold 
$\Sigma$ is known to be  
an infinite dimensional 
Riemannian manifold. Coordinates are known which cover it and the 
tangent space, metric, connection and curvature tensor are known. So 
the issue is not just the infinite dimensionality.  The problem is 
that the diffeomorphism group is very large and its action is quite
complicated.  It is known that the quotient has singularities and 
is not everywhere an infinite dimensional manifold. 

To define the quotient we must also be able to have a procedure
to answer the following question:
given two metrics, $g_{ab}$ and $g_{ab}^\prime$ on $\Sigma$, does there
exist a diffeomorphism $\phi \in Diff(\Sigma )$ such that 
$g_{ab}^\prime = \phi \circ g_{ab}$? By computing a large enough set 
of curvature scalars one can reduce this problem to the question of 
when there is a diffeomorphism that can bring one set of functions on 
a manifold to another.  But 
no finite procedure is known
which can answer this question in general even in the case that we
restrict to the analytic category in which both metrics and 
diffeomorphisms can be defined by power series expansions in some set 
of coordinates.  The problem is that while one may be able to show in 
a finite number of steps that $g_{ab}$ and $g_{ab}^\prime$
are not diffeomorphic, there is no finite procedure which works in the
general case to tell whether 
they are or not.

But if one cannot tell whether two metrics are in the same 
diffeomorphism class or not one is not going to be able to  
construct the quotient by an explicit construction of the equivalence 
class within the space of metrics on $\Sigma$.  If this is the case, postulate
{\bf D} is not satisfied\footnote{Note that if the exact 
configuration space is not constrictible, one may still {\it postulate}
coarse grained configuration spaces such as mini-superspace models.
However, these are then not strictly speaking derivable from general
relativity, while they may be suggested by some heuristic 
considerations based on general relativity, 
they must be considered to be logically independent of
full general relativity.}.

One might try to avoid this by use of the alternative Postulate
${\bf D^\prime}$, which replaces the requirement of constructing the space of 
metrics mod diffeomorphisms by the solution of certain functional 
differential equations.  This is suggested by the fact that in 
one approach to quantum gravity a complete solution to the
problem of solving the diffeomorphism constraints has been stated
in closed form.  This is the loop quantum gravity approach\cite{loops}.  
The basic result of loop quantum gravity is a precise recipe for the
construction of an orthonormal 
basis for the Hilbert space of quantum general relativity on a 
spatial manifold with topology and differential structure $\Sigma$. 
The basis is in one to one correspondence with the diffeomorphism 
equivalence classes of the embeddings in $\Sigma$ of a certain class 
of labeled graphs called spin networks. 
The problem we are concerned with then reduces to the question of
classifying the diffeomorphism classes of embeddings of these graphs.

There is no problem enumerating and distinguishing finite labeled 
graphs as this comes down to distinguishing finite dimensional 
matrices with integer entries.  However, classifying and 
distinguishing diffeomorphism classes of the embeddings of graphs in 
three dimensions is a tricky question.  The problem of the 
classification of ordinary knots up to ambient 
isotopy\footnote{Ambient isotopy is a weaker equivalence, which
requires only that the embeddings of the knots or graphs are 
homeomorphic. This does not, in particular, preserve 
differentiability at nodes of the graphs.} has only 
recently been solved and even that requires a very computationally 
intensive procedure which involves classifying the finite groups 
that appear as the homotopy group of the complement of the knot.
The homotopy groups of complements of graphs present a greater 
challenge, and, to my knowledge, these remain unclassified.

Even if this problem is solved the problem of distinguishing 
diffeomorphism classes is a good deal trickier for graphs with 
arbitrary valence intersections than the problem of distinguishing 
ambient isotopy classes.  The reason is that in the case of 
sufficiently high valence ($5$ and higher in three spatial dimensions)
the diffeomorphism invariance classes are labeled by continuous 
parameters.

If it turns out that there is no finite procedure for solving either 
problem then it 
follows there will be no finite procedure to construct the Hilbert 
space for quantum general relativity and related theories\footnote{including 
all known couplings to matter including supersymmetric theories.}.

The problem is made even more complicated if one believes, as many do, 
that the topology of $\Sigma$ should also be able to change by means 
of quantum transitions.  In this case one has to 
classify networks embedded in arbitrarily complex three topologies.
But it is sufficient to note that the problem 
is already quite serious for a fixed simple topology.

Finally, we note that even if the answer to the first question is no, 
the answer to the second is certainly yes. The  
maximum computational power of a universe could not increase
faster than its volume, which is roughly proportional to the number of nodes of 
the graph of the corresponding quantum state.  But the number of steps 
required to distinguish the embeddings of two graphs is likely to go up 
faster than any power of the number of nodes.  We may also note that 
the situation is likely worse than this if the holographic 
principle\cite{holo1,holo2,louis-holo,weakholo,weakstrong}
is correct, as that bounds the amount of information a region of 
space could contain to its area rather than its volume in Planck units.

To summarize this part of the argument: it is quite possible that the 
answer to the first question raised above is yes, and almost certain 
that the second is.  In the case that the configuration space cannot 
be constructed the argument for the elimination of time in classical 
cosmology cannot be run, for the whole framework falls apart without a 
pre-specified configuration space.  Similarly, if the Hilbert space of states 
cannot be constructed the argument for the elimination of time in 
quantum cosmology cannot be run.

In either case, if the spaces are constructible, but require more 
computational power than the universe could contain, then we are 
faced with an interesting situation.  A mythical extra-universe 
observer could run the argument for the elimination of time.  But this
is impossible for any real observers inside the universe. A quantum 
theory of cosmology that requires more processing power than the 
universe could contain to set up its Hilbert space and check whether 
two states were orthogonal or not would not be a theory that we who 
live inside the universe could use to do real computations.  So it 
is not clear what relevance for our physics there may be for the 
possibility that some imagined being outside the universe could 
eliminate time from physics. What matters is that we cannot work with 
a physical theory without time.

\section{Conclusions}

Let me begin my summary of these arguments by emphasizing the role 
played by the requirement that a theory of cosmology must be 
falsifiable in the usual way that ordinary classical and quantum 
theories are.  This leads to the requirement that  a sufficient 
number of observables can be determined by information that reaches a 
real observer inside the universe to determine either the classical 
history or quantum state of the universe.    Only if this is the 
case can we do cosmology within the standard ideas concerning the 
methodology and epistemology of dynamical theories.  

It does appear that this is not the case.  Interestingly enough, this 
statement requires input from both theory and observation.  General 
relativity allows both possibilities.  Presently observations seem to 
rule out the possibility that we live in a universe in which there is 
a complete Cauchy surface within the classical region of the 
past light cones of typically 
situated observers such as ourselves.  

The implication seems clear: if we want to do cosmology as a science, 
we must restrict ourselves to theories in which all observables are 
accessible to real observers inside the universe.  To do this we must 
invent a new framework for quantum cosmology which does not include 
notions like ``the wavefunctional of the universe."  One way to do this 
has been proposed by Markopoulou, which is called ``quantum causal 
histories"\cite{FM4,FM5}.   These are  an 
interconnected web of Hilbert spaces tied to the causal structure in 
such a way that each act of observation, considered as a particular 
event in the history of the universe, is represented in terms of a 
Hilbert space constructed to represent the information available to be 
observed at that event.  These together provide a representation of 
projection operators whose 
spectrum consists of the observer dependent truth values 
we discussed above.

This proposal may be seen in the light of a general issue which has 
been discussed a great deal in quantum cosmology, which is that of 
context dependence.  This arises most generally in the consistent or 
decoherent history approaches to quantum 
cosmology\cite{consistent,GMH}.  
As pointed out 
by Gell-Mann and Hartle\cite{GMH} and Dowker and Kent\cite{DK}, 
the consistent histories 
formulation requires the specification of a context within which 
observations are to be made, prior to their interpretation.   This is
necessary to replace the quantum world/classical world division of
Bohr's interpretation of quantum theory in a way that avoids the 
preferred basis problem of the Everett interpretation.  It may also
be argued that the consistent histories approach does evade many
of the issues which face a hamiltonian approach to quantum cosmology,
including those discussed here\footnote{I would like to thank James 
Hartle for correspondence on this point.} 
Isham and 
collaborators\cite{chris-topos} have studied the general 
mathematical structure of such 
contextual approaches to cosmology and found they are naturally 
formulated in terms of topos theory.  Markopoulou's quantum casual 
histories\cite{FM4,FM5} can be understood from this point 
of view as the result of 
using the causal structure of the history of the universe to define 
the contexts.  

However, the general issue of context dependence does not by itself 
refute the argument for the elimination of time.  
Were it possible for a single observer inside the universe 
to make sufficient measurements to determine either its classical 
history or its quantum state, and assuming that the other four
postulates also held, the argument for the elimination of time 
could be run.  In this circumstance it might still be convenient to 
express quantum cosmology in terms of histories, but it would not be 
essential.  Barbour and others would be able to argue that they could 
do cosmology perfectly well with no notion of histories apart from 
what was necessary to recover the classical limit.  

It is only  by insisting that the context of real observers 
inside their universe is defined by the information that reaches them 
by means of radiation that propagated from their past that a link is 
made between the issue of observability of the universe and its causal 
structure.  Of course,  the notion of causal structure may be loosened quite 
a bit from that which arises in general relativity, as has been done 
in various causal set and evolving spin network models of quantum 
gravity.  But it is hard to divorce the notion of causal structure 
from the idea that there is a finite speed for the propagation of 
information, and hence from some notion of time.  

Is the notion of time then built into the argument from the beginning?
No, the key point is the insistence on building a cosmological theory 
that makes references only only to observations made by real 
observers inside the universe.  It is then an observed fact that the 
universe is very big compared to its observers. A combination 
of observation and theory then leads to the conclusion that the 
observations made by one observer at one ``moment" are insufficient to 
determine the classical or quantum state of the whole universe.  
We may note that all that is required here is the notion of a moment
at which a number of simultaneous measurements may be made.
This
is already all we need to argue that cosmology requires a different 
framework from a conventional quantum system because the postulates of 
quantum theory require that the state of a quantum system must be 
uniquely determined by measurements of a complete set of commuting 
observables which, by definition, can be made simultaneously.

This is already sufficient to refute the argument for the elimination of 
time. The proposal of Markopoulou builds up from here to propose an 
alternative way to do cosmology, which is in terms of a structured set 
of observations, which are not made at simultaneous moments.  The 
positive proposal is that the structure which is imposed on the 
possible measurements is a partial ordering which is derived from the 
causal structure of the universe.   

This is a good moment to recall Newman's worry.  This may be put in 
the following form. Even if the universe is governed by classical 
deterministic equations, these are likely to be chaotic. This 
is important because of the fact that all 
measurements we real observers can do are of finite precision. It 
means that, even if all five postulates are true, and 
the argument for the elimination of time can be run, we still may
not be able to make reliable predictions of measurements that 
refer to ``moments" when our clocks read different ``times."  Even if 
time can be eliminated in principle, it cannot be eliminated in 
practice.  

There is a stronger form of Newman's worry, which has to do with the 
existence of a physical observable algebra for a chaotic but time 
reparameterization invariant system.   But even leaving this aside we 
see that this argument has an interesting parallel with Markopoulou's 
argument. Both refer to the information available to a real observer 
inside the universe.  They concern two ways in which real observers 
differ from abstract idealized observers in the real universe.  
First, that they are small and are limited to information available in 
a local region of spacetime. Second, they can only make observations 
of finite precision.  It is very interesting that both arguments 
suggest it may be useful to think about measurement in information 
theoretic terms.  In particular both  suggest it is important to make
a distinction between what can be done in principle in a mathematical 
model and what can be done by a real observers in a finite number of steps.  

A related issue of constructibility underlies the third argument, 
introduced by Kauffman.  The implication of this argument is that if 
we want to have a measurement theory which is relevant to what real 
observers inside the universe do we must require that all the 
constructions necessary to formulate and compute in the theory 
can be carried out in a number of operations which is not only finite 
but small enough to be carried out by an observer inside the universe.

In the cases of theoretical biology and economics, Kauffman goes on to 
propose a method for formulating a theory that does not depend on the 
constructibility of huge and complex configuration spaces.  He 
proposes that it may be sufficient that the theory provide an 
algorithm that allows the construction of all configurations which 
differ by a small number of local changes from any given present 
configuration.  This space of possible nearby configurations is large 
enough for a large complex system, but still infinitely smaller than 
the space of all possible configurations.   Kauffman calls it {\it the 
adjacent possible\cite{stu-invest}.}

Could we do quantum cosmology in terms of such an adjacent possible?
A start on such an approach has been given in \cite{stulee2}.
Suppose that the state space of a quantum theory of gravity was not 
constructible.  For example suppose this is in fact the case for the 
ambient isotopy classes of embeddings of spin networks in a three 
manifold.  It is still possible to construct the space of all spin 
networks that differ from any given one, $\Gamma$, by a 
small number of  local moves.  The local moves can be moves that 
involve rearrangements of less than $P$ nodes, which together with 
their common edges must make an $n$ simplex for $n \leq P$.  These
are the forms of moves appropriate to $P$ spacetime dimensions.  
Given $\Gamma$ let $\Omega_\Gamma^r$ be the linear space spanned by 
all spin networks that are the result of $r$ local moves made on 
$\Gamma$.  This will be a finite dimensional space.  

We may now use 
the following facts from the theory of causally evolving spin 
networks\cite{FM2}: 1) The evolution moves are 
themselves local moves of the same type. 2) Each local change 
constitutes the analog of an event in a quantum 
spacetime.  3) Each such event will 
correspond in the classical limit to a small, but finite spacetime.
We can then deduce that the space of quantum universes which 
begin with the initial state $\Gamma$ and have finite spacetime 
volume, $V$, in Planck units, 
will live in some $\Omega_\Gamma^r$, with $r$ a finite
function of $V$.  But for any finite $r$ and finite $\Gamma$, 
$\Omega_\Gamma^r$ is a finite dimensional space.
Using this kind of construction it does appear that quantum
cosmology could be formulated in terms of an adjacent possible
type construction\cite{stulee2}.

However, the resulting theories are still subject to the argument of 
Markopoulou: the states in $\Omega_\Gamma^r$ are not measurable by any 
observer inside the universe.  To take into account both we should 
realize this kind of construction in a quantum causal history.
This is in any case a natural thing to do as a quantum causal history 
is based on a partial order structure, and such a structure is naturally 
generated by local moves made on graphs.  The result is a framework 
for quantum cosmology which escapes the worries raised here.

In these theories the notion of information plays a crucial role.  
This is forced on us by the combination of
finiteness of the space of local changes and finiteness of the
propagation of the effects of local changes.  But the importance
of the notion of information also arises from an 
apparently independent set of considerations having to do with the 
Bekenstein bound\cite{bek} and the 
holographic principle\cite{holo1,holo2,louis-holo,weakholo,weakstrong}. 
These suggest that 
measures of geometry may actually be reduced to measures of 
information flow. In its "weak form", the holographic principle
asserts that the geometric area of any surface must be reducible,
in a fundamental theory, to a measure of the capacity of that surface 
as a channel of flow of information from its causal past to its causal 
future\cite{weakholo,weakstrong}.    

In closing we note the very interesting way in which notions of 
finiteness and constructibility are coming into fundamental theories 
of quantum cosmology.  This may not be surprising to experts in 
philosophy and the 
foundations of mathematics, for whom these notions are closely related 
to ideas on time.  But it is a new idea for some of us who come at it 
from physics: if the universe is discrete and time is real, and 
is itself composed of discrete steps, then time 
may be none other than the process which constructs, not
only the 
universe, but the space of possible universes relevant for observations
made by local observers.  Beyond this, there is the 
possibility of a quantum cosmology in which the actual history of 
the universe up till some moment and the space of possible universes 
present at that ``instant"
are not two different things, but are just different ways of seeing the 
same structure, whose construction is the real story of the world.

\section*{ACKNOWLEDGEMENTS}

This paper was mainly a commentary on ideas of Barbour, Kauffman, Markopoulou 
and Newman, and I would like to thank them for many discussions on 
these issues.  I am also grateful
to James Hartle for commenting on a draft of this paper and Chris 
Isham and Carlo 
Rovelli for discussions on these issues.  I would like also to thank the
theoretical physics
group at Imperial college for their hospitality during this last
year. This work is supported by the
NSF through
grant PHY95-14240 and a gift from the Jesse Phillips
Foundation.

\end{document}